# Measuring Affectiveness and Effectiveness in Software Systems

Giuseppe Destefanis[2], Marco Ortu[1], Steve Counsell[2], Michele Marchesi[1], Roberto Tonelli[1]

[1]DIEE, University of Cagliari
{marco.ortu,michele,roberto.tonelli}@diee.unica.it

[2]Brunel University, London
{giuseppe.destefanis,steve.counsell}@brunel.ac.uk

*Abstract*—The summary presented in this paper highlights the results obtained in a four-years project aiming at analyzing the development process of software artifacts from two points of view: Effectiveness and Affectiveness.

The first attribute is meant to analyze the productivity of the Open Source Communities by measuring the time required to resolve an issue, while the latter provides a novel approach for studying the development process by analyzing the affectiveness ex-pressed by developers in their comments posted during the issue resolution phase.

Affectivenes is obtained by measuring Sentiment, Politeness and Emotions. All the study presented in this summary are based on Jira, one of the most used software repositories.

## I. INTRODUCTION

Software Engineering has many goals, among them we can certainly consider monitoring and controlling the development process in order to meet the business requirements of the final software artefact and to guarantee its quality [7], [10], [11]. During the development phases, software engineers need to have empirical evidence that the development process and the overall quality of software artifacts is converging to the required features. Improving the development process's Effectiveness leads to higher productivity, meaning shorter time to market, but given the immateriality of software, understanding or measuring software development is a very hard challenge. Modern software is the result of a complex process involving many stakeholders such as product owners, quality assurance teams, project managers and, above all, developers. All these stakeholders use complex systems like issue tracking, code versioning and release scheduling.

In Open Source Development the situation is even more complicated given the structure of the open source communities, often spread around the globe, with different time shifts, cultures, languages and environments. Beside their complexity, Open Source Communities provide valuable empirical data through source code repositories, issue tracking systems, mailing list, which can help researchers in helping developers and managers on how to improve software quality.

Tools for project management and issues/bugs tracking are becoming useful for governing the development process of Open Source software, because they are able to simplify the communications process among developers and ensure scalability of a project. The more information developers are able to exchange, the clearer the goals become.

By analyzing data stored in such systems, researchers are able to study and address questions such as: what are the factors able to impact/improve productivity? Is it possible to improve software productivity shortening the time to market?

The present work summarizes the main results obtained by the authors at the end of the project, with the main goal of building a bridge between researchers and developers, in order to provide (hopefully) useful information able to help developers in their very complex and high demanding daily job.

## II. RELATED WORK

Research has focused on understanding emotions and mood both in software engineering and software development and how the human aspects of a technical discipline can affect final results [4],[5],[6],[12], [20],[21],[23]. Several studies have also investigated the relationship between affect and work-related achievements, including performance [22] and problem-solving processes, such as creativity [2].

In this paragraph we provide an overview of the main related works in the field of behavioural software engineering and emotions, which inspired the authors during the 4 years-project, indicating the year of the study throughout the discussion.

In 2007, Rigby et al. [31] analyzed, using a psychometrically-based linguistic analysis tool, the five big personality traits of software developers in the Apache httpd server mailing list. The authors found that two developers responsible for the major Apache releases had similar personalities and their personalities were different from other developers.

In 2008, Acuna et al. [1], performed empirical research examining the work climate within software development teams. The authors attempted to understand if team climate (defined as the shared perceptions of team work procedures and practices) bore any relation to software product quality. They found that high team vision preferences and high participative safety perceptions of the team were significantly related to better software. Feldt et al. [13] focused on personality as a relevant psychometric factor and presented results from an empirical study about correlations between personality and attitudes to software engineering processes and tools. The authors found that higher levels of the personality dimension "conscientiousness" correlated with attitudes towards work style, openness to changes and task preference.

In 2013, Bazelli et al. [3] analyzed questions and answers on stackoverflow.com to determine the developer personality traits, using the Linguistic Inquiry and Word Count [30]. The authors found that the top reputed authors were more extroverted and expressed less negative emotions than authors of down voted posts. Guzman et al. [18], [16] have proposed prototypes and initial descriptive studies towards the visualization of affect over a software development pro-cess. In their work, the authors applied sentiment analysis to data coming from mailing lists, web pages, and other text-based documents of software projects. Guzman et al. built a prototype to display a visualization of the affect of a development team, and they interviewed project members to validate the usefulness of their approach. In another study, in 2014, Guzman et al. [17], performed sentiment analysis of Github's commit comments to investigate how emotions were related to a project's programming language, the commits' day of the week and time, and the approval of the projects. The analysis was performed over 29 top-starred Github repositories implemented in 14 different programming languages. The results showed Java to be the programming language most associated with negative affect. No correlation was found between the number of Github stars and the affect of the commit messages.

In 2014, Graziotin et al. [14] reported the results of an investigation with 42 participants about the relationship between the affective states, creativity, and analytical problem-solving skills of software developers. The results offered support for the claim that happy developers were better problem solvers in terms of their analytical abilities. The authors provided a better understanding of the impact of affective states on the creativity and analytical problem-solving capacities of developers, introduced and validated psychological measurements, theories, and concepts of affective states, creativity, and analytical-problem-solving skills in empirical software engineering, and raised the need for studying the human factors of software engineering by employing a multi-disciplinary viewpoint.

In another work, Graziotin et al. [15] conducted a qualitative interpretive study based on face-to-face open-ended interviews, in-field observations and e-mail exchanges. This enabled the authors to construct a novel explanatory theory of the impact of affects on development performance. The theory was explicated using an established taxonomy framework. The proposed theory built upon the concepts of events, affects, attractors, focus, goals, and performance.

## III. CONNECTING THE DOTS

Our journey started when we found that comments posted by developers on software repositories contain not only technical information, but also valuable information about sentiments and emotions. We then started the creation of a dataset, now publicly available [28], [29], hosting more than 1K projects, 700K issue reports and 2 million comments. We fetched the data by mining the Jira repository of four open source communities: Apache, Spring, JBoss and CodeHaus and we also presented the tools used for the mining activity and how information is organized in the dataset. We manually labeled 2,000 issue comments and 4,000 sentences written by developers with emotions such as love, joy, surprise, anger, sadness and fear. By sharing the repository, we wanted to fill the gap of missing data in the area and to encourage the research community to perform studies in the field of software emotions.

The second step has been the study of the developer networks of the open source projects hosted in JIRA [27]. By analyzing 7 big open source projects, we further investigated how the productivity was distributed across the communities. To measure the productivity we considered factors such as the community size, the number of fixed issues, the distribution of fixed issue's maintenance type and priority, and the average issue fixing time. As a first result we found the presence of Pareto's law (20% of developers doing 80% of the work), and that there were a few developers that posted and commentated the majority of issues. We showed the independence of the average issue resolution time from the other factor considered, such as the community size and the kind of issues maintenance and priority. There are many other factors that may impact the average community issue fixing time, for example software component involved in the issue resolution or the portion of code involved. This result agrees with other studies about the social structure of open source projects. We further investigated how the productivity was distributed across the communities.

As a third step we focused our attention in understanding the relationship among sentiment, emotions, politeness and productivity measured in terms of "issue fixing time". In [26], [8] we presented the results about politeness and attractiveness on 22 open-source software projects developed using the Agile board of the JIRA repository. Our results showed that the level of politeness in the communication process among developers does have an effect on both the time required to fix issues (Fig. 2) and the attractiveness of the project to both active and potential developers. The more polite developers were, the less time it took to fix an issue. In the majority of cases, the more the developers wanted to be part of project, the more they were willing to continue working on the project over time.

In [24] we showed that the three affective metrics, i.e., emotions, sentiment and politeness, were independent, showing a weak correlation of at most 0.36, in contrast to some of the control metrics who obtained a moderate to strong correlation among themselves of at most 0.7. Then, we showed how affectiveness metrics statistically improve an explanation model of issue fixing time compared to a model based on control metrics. The 4th, 5th and 6th most important metrics in the model corresponded to % of love comments (-50.19%), issue average politeness (+49.76%) and % of sadness comments (+38.39%). In other words, comments containing JOY and LOVE emotions had shorter issue fixing time, while comments containing SADNESS emotion had a longer fixing time. Although we found that the politeness of the last comment has a shorter issue fixing time, it was unexpected that less polite comments were linked with shorter fixing time. After investigation we found that for about the 50% issue reports with extreme politeness (polite and impolite) had shorter issue fixing time. Those reports tended to only have a median number of 2 developers discussing the issue, and the negative issues had the lowest number of sentences in the comments.

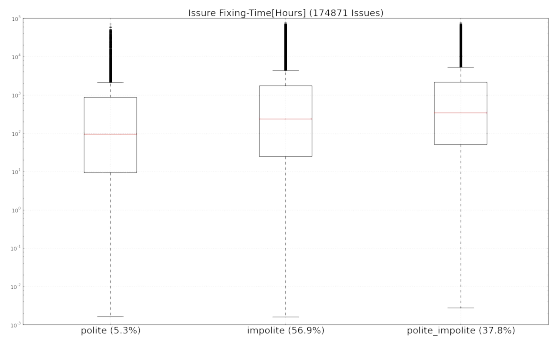

Fig. 1: Box-plot of the fixing-time expressed in Hours. The number in parentheses next to issue group indicates the percentage of issues

The results about correlation among affective metrics have been deeply studied in [9], in which we presented correlation results related to seasonality and randomness of sentiment and emotions time series from 10 open-source software projects. The results showed that there was not significant correlation among *joy and love, sadness and anger, sentiment and joy*; each time series therefore brings different information. We found that *joy, love, sadness and anger*, for the majority of the projects in our corpus were seasonal and not random. This is an interesting fact that could help managers and developers in better understanding the development process and in managing the all the activities trying to resolve conflict and avoid negative emotions which could affect the productivity of the developers involved in the development process and the final quality of the system being developed

We concluded our studies with [25] in which we empirically determined how developers interacted with each other under certain psychological conditions generated by politeness, sentiment and emotions of a comment posted on Jira. Results showed that when in the presence of impolite or negative comments, there was higher probability for the next comment to be neutral or polite (neutral or positive in case of sentiment) than impolite or negative. This fact demonstrates that developers, in the dataset considered for the study, tended to resolve conflicts instead of increasing negativity within the communication flow. This is not true when we considered emotions; negative emotions are more likely to be followed by negative emotions than positive. Markov models provide a mathematical description of developer behavioural aspects and the result could help managers take control the development phases of a system (especially in a distributed environment), since social aspects can seriously affect a developer's productivity.

## IV. THREATS TO VALIDITY

Several threats to validity could affect the results of the findings of the different studies summarized in this paper.

We supposed, based on empirical evidence, a relationship between the emotional state of developers and what they write in issue reports. Since the main goal of developer communication is the sharing of information, the consequence of removing or camouflaging emotions may make comments less meaningful and cause misunderstanding.

All the works were focused on sentences written by developers for developers. To illustrate the influence of these comments, it is important to understand the language used by developers. The comments used in the studies were collected over an extended period from developers unaware of being monitored. For this reason, we are confident that the emotions we analyzed were genuine.

The politeness measures were approximated and could not perfectly identify the precise context, given the challenges of natural language and subtle phenomena like sarcasm. Another major threat has been highlighted in a recent study by Jongeling and al. [19], in which the authors studied the impact of the choice of a sentiment analysis tool when conducting software engineering studies. The authors observed that the tools considered do not agree with the manual labelling, but also they do not agree with each other, that this disagreement can lead to diverging conclusions and that previously published results (included several studies presented in this summary) cannot be replicated when different sentiment analysis tools are used. The results suggested a need for sentiment analysis tools specially targeting the software engineering domain.

## V. CONLCUSIONS

Human Affectiveness such as the emotional state of a person influences human behaviour and interaction. Software development is a collaborative activity and thus it is not exempt from such influence. Affective analysis, e.g., measuring emotions, sentiment and politeness, applied to developer issue reports, can be useful to identify and monitor the mood of the development team, allowing project leaders to anticipate and resolve potential threats to productivity (especially in remote team settings), as well as to discover and promote factors that bring serenity and productivity in the community.

This paper presented a summary which highlights the main results obtained by the authors during a 4-years project. Above all, the takeaway message of all the studies presented is that positive emotions and good manners are good for both productivity and wellness of developers.